\documentclass[aps,prd,reprint,superscriptaddress,longbibliography,nofootinbib]{revtex4-2}
\usepackage{amsmath,amssymb,bm,mathrsfs,graphicx}
\usepackage{pgfplots}
\pgfplotsset{compat=1.18}
\usepackage[hidelinks]{hyperref}

\begin{document}
\title{The fate of primordial gravitational waves during domain-wall domination}
\author{Zhen-Min Zeng}
\affiliation{Institute for Theoretical and Mathematical Physics, MSU, 119991 Moscow, Russia}
\begin{abstract}
Primordial gravitational waves probe inflation, but their interpretation
 depends on their subsequent evolution. During domain-wall domination,
 a frozen wall network retains a nonzero shear response that can modify
 the superhorizon evolution of primordial gravitational waves. Starting
 from the wall-forming scalar field,
 we derive an effective tensor mass term and a source associated with the
 frozen wall geometry in the thin-wall, quasistatic limit. Together they drive superhorizon tensor perturbations
 toward a configuration set by the network. This configuration carries both any inherited primordial
 deformation and the network's intrinsic random structure. We derive how
 these contributions enter the tensor transfer and power spectrum. The
 resulting description connects superhorizon tensor evolution to the
 microscopic wall response and the geometry established as the network
 freezes.
\end{abstract}
\maketitle

\section{Introduction}
\label{sec:introduction}

Primordial gravitational waves carry information about inflation,
including its energy scale and the dynamics of tensor
production~\cite{Starobinsky:1979ty,Lyth1997,Stewart:1993bc,Guzzetti2016}. Extracting this information requires
understanding how the primordial signal evolves through the subsequent
cosmic history. Radiation domination is well supported during primordial
nucleosynthesis, but the earlier expansion history is much less
constrained, allowing nonstandard epochs that can affect the primordial
gravitational-wave signal~\cite{Mendes_1999,Giovannini_1998,Seto:2003kc,Tashiro_2004,Boyle_2008,Allahverdi2021}.
Here we consider a transient period of domain-wall domination (DWD).

Such a period is possible because
walls dilute more slowly than radiation. For approximately constant
tension $\sigma$, radiation-era scaling gives
$\rho_{\rm DW}\sim\sigma/t$, whereas radiation dilutes as $t^{-2}$~\cite{Kibble1976,Zeldovich:1974uw,Press:1989yh,Kawano:1989mw,Hindmarsh_1996,Saikawa2017}.
The walls can therefore dominate before decaying and restoring radiation
domination before primordial nucleosynthesis~\cite{Kawasaki_2005,Hattori:2015xla,de_Salas_2015,Hong2026}.
Reference~\cite{Zeng2026} demonstrated that domain-wall networks
approach a dynamically frozen state as they come to dominate the Universe,
with vanishing velocities and an asymptotically fixed comoving configuration.
This freezing suggests that the wall-dominated Universe may behave as a relativistic solid, which has non-vanishing shear response~\cite{Bucher1999,Battye2005,Battye_2005}.

Can primordial gravitational waves evolve outside the horizon during
this frozen DWD epoch? Addressing this
question therefore requires understanding the tensor stress of the
frozen network. In this work, we derive the effective tensor mass term and the
frozen-network source from first principles, starting from the stress
tensor of the wall-forming scalar field. In the thin-wall limit, we
include the quasistatic adjustment of the local wall profile to a tensor
deformation and average the stress over wall positions and orientations.
The metric-induced response recovers the known shear modulus of an
isotropic wall ensemble~\cite{Battye2005,Battye_2007} and yields the effective tensor
mass term, while the configuration-dependent stress supplies the
frozen-network source.

Using this description, we find that superhorizon tensor perturbations
relax toward a configuration determined by the frozen network.
The surviving primordial correlation depends on how much of the
primordial deformation the network inherits. For example, adiabatic
inheritance preserves the primordial constant component, while the
absence of inheritance leads to its decay. An effective tensor mass
therefore does not by itself determine the primordial tensor transfer. We quantify this dependence separately from the additional
tensor power sourced by the network's intrinsic random structure.
The inheritance realized by a physical network must be determined
from its formation and freezing dynamics.

Our calculation applies to modes that remain superhorizon throughout
the ideal frozen DWD interval and have wavelengths much larger than the
comoving network correlation length. We first specify the frozen
background, then derive the effective tensor mass and source, and
calculate the tensor evolution and power. Appendix~\ref{app:covariant}
gives the covariant thin-wall stress and its variation before the
fixed-surface limit is imposed.

\section{Domain-wall domination and dynamical freezing}\label{sec:background}
The dynamical basis for the frozen network was established in our
previous work~\cite{Zeng2026}, which coupled the velocity-dependent
one-scale model~\cite{PhysRevD.72.083506} self-consistently to the Friedmann equation and
radiation energy transfer. In that system, gravitational backreaction
destabilizes the radiation-era scaling solution, and the evolution
approaches a wall-dominated regime. Writing $v_{\rm rms}$ for the
root-mean-square physical wall speed, $L_{\rm phys}$ for the physical
network correlation length, and $a$ for the scale factor, this limit is
\begin{equation}
 v_{\rm rms}\longrightarrow0,\qquad
 \frac{L_{\rm phys}}{a}\longrightarrow L_{\rm eq}
 =\mathrm{const}.
 \label{eq:freezing}
\end{equation}
Here $L_{\rm eq}$ is the limiting comoving correlation length.
The walls thus become asymptotically frozen on comoving surfaces.
In this model, freezing follows from the coupled network--expansion
dynamics; we calculate the tensor response about this limiting state.

In the limiting state, $L_{\rm phys}\propto a$,
$\rho_{\rm DW}\propto a^{-1}$, and the ratio of mean wall pressure
to energy density is $w=-2/3$. We approximate the approach
to this state using conformal time $\tau$, defined by $dt=a\,d\tau$
for cosmic time $t$. The ideal frozen, pure DWD epoch starts at
$\tau_{\rm eq}$ and ends at $\tau_{\rm ann}$, when annihilation begins to invalidate
the frozen description.
The idealization does not assert instantaneous freezing in the physical
mixed radiation--wall background at equality. During ideal DWD,
$a\propto(\tau_*-\tau)^{-2}$,
where $\tau_*$ is the future conformal boundary of the ideal DWD
solution, $\mathcal H=a'/a=aH$, and a prime denotes $d/d\tau$.
Throughout, $\mathcal H$ denotes the conformal Hubble parameter and
$H$ the physical Hubble rate. The finite transition to freezing remains a matching uncertainty.

\section{Scalar-field origin of the effective tensor mass}\label{sec:mass}
We derive the metric-induced anisotropic stress directly from the
scalar field and identify its contribution to the effective tensor mass.
For the canonical wall-forming scalar $\phi$ with potential $V(\phi)$,
the mixed spatial stress is
\begin{equation}
 T^i{}_j=g^{ik}\partial_k\phi\partial_j\phi
 -\delta^i{}_j\left[\frac12g^{\alpha\beta}
 \partial_\alpha\phi\partial_\beta\phi+V(\phi)\right].
 \label{eq:scalarstress}
\end{equation}
Here $g^{\mu\nu}$ is the inverse spacetime metric; Greek indices
span spacetime and Latin indices space. The Kronecker term
$\delta^i{}_j$ is purely isotropic and has
zero transverse-traceless (TT) projection; the tensor source comes from the first term,
including its metric and field-profile dependence.
We use a spatially flat
Friedmann--Robertson--Walker (FRW) background $ds^2=a^2[-d\tau^2+(\delta_{ij}+h_{ij})dx^i dx^j]$,
where $x^i$ are comoving coordinates and $h_{ij}$ is the dimensionless
tensor metric perturbation, with $\partial_i h_{ij}=h_{ii}=0$ and
$g^{ij}=a^{-2}(\delta^{ij}-h^{ij})+O(h^2)$.
Spatial indices on perturbations are raised with $\delta^{ij}$.
For Fourier amplitude $h_A(\bm k,\tau)$, with $k=|\bm k|$ and
polarization $A=+,\times$, the tensor Einstein equation is
\begin{equation}
 h_A''+2\mathcal Hh_A'+k^2h_A=16\pi G a^2 \bigl[T^i{}_j\bigr]_A^{\rm TT},
 \label{eq:einstein}
\end{equation}
Here $[T^i{}_j]_A^{\rm TT}$ denotes the Fourier-space TT projection
of the mixed spatial stress onto polarization $A$.
Crucially, the stress depends on the tensor metric itself.
For example, the spatial-gradient contraction in
Eq.~\eqref{eq:scalarstress} contains
\begin{equation}
 g^{ij}\partial_i\phi\partial_j\phi
 =a^{-2}\left[(\bm\nabla\phi)^2
 -h^{ij}\partial_i\phi\partial_j\phi\right]+O(h^2).
 \label{eq:coupling}
\end{equation}
Here $\bm\nabla$ is the coordinate spatial gradient.
The wall gradients belong to the microscopic background, so the term
$h^{ij}\partial_i\phi\partial_j\phi$ is linear in the tensor
perturbation. The associated stress response is of order
$\rho_{\rm DW}h$, where $h$ denotes a characteristic tensor amplitude.
During DWD, the Friedmann equation gives
$Ga^2\rho_{\rm DW}\sim\mathcal H^2$, so this response contributes
at order $\mathcal H^2h$ to Eq.~\eqref{eq:einstein}.
It therefore competes with Hubble evolution and must be retained
even on superhorizon scales.
To obtain the TT stress response, we vary both the explicit metric
dependence of $g^{ik}\partial_k\phi\partial_j\phi$ and the wall profile.

\begin{figure*}[t]
 \centering
 \includegraphics[width=0.82\textwidth]{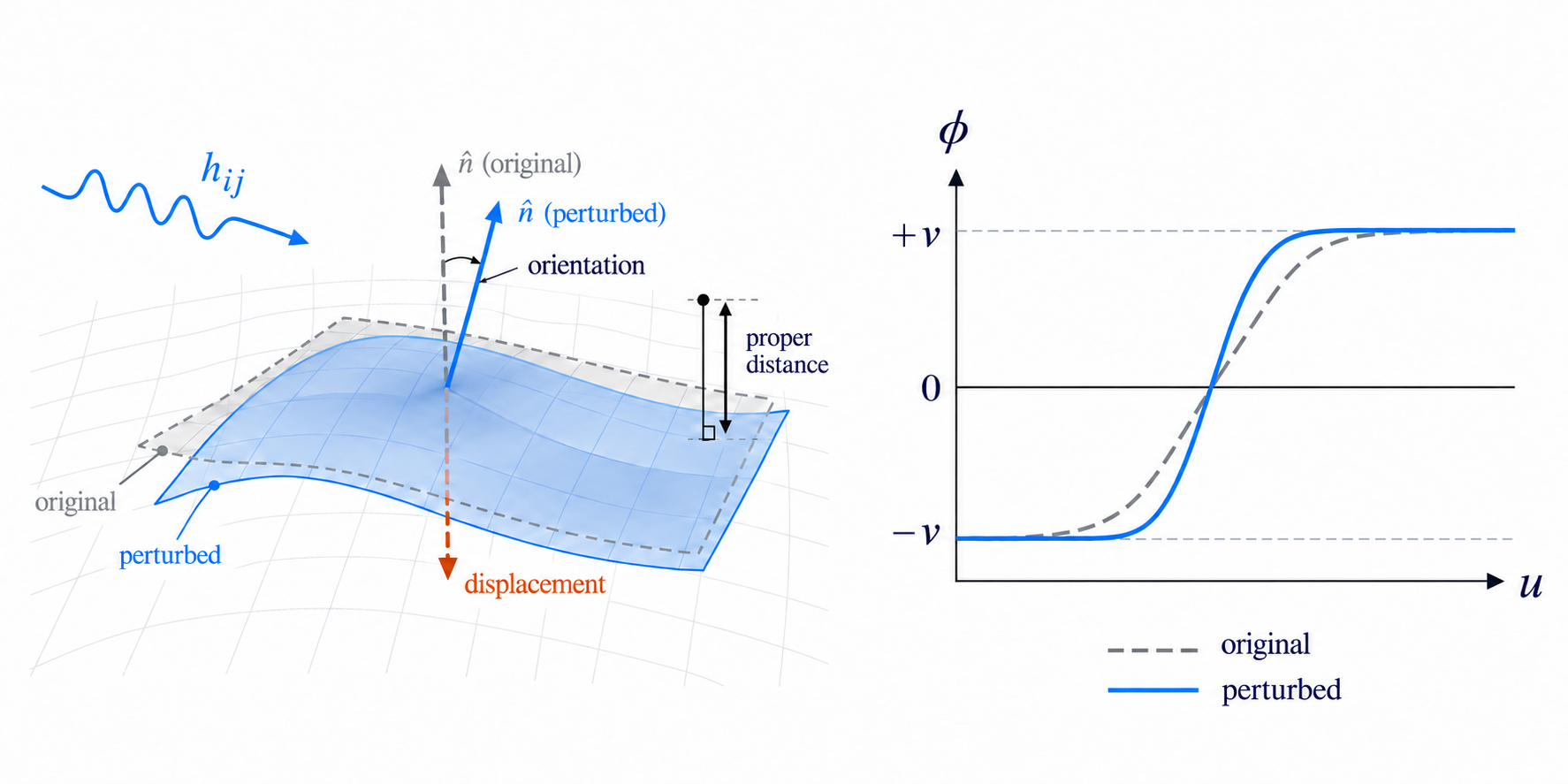}
 \caption{Schematic tensor--wall response. Left: changes in wall
 orientation and normal proper distance, with a possible displacement.
 Gray dashed and blue surfaces denote the original and perturbed walls.
 Right: the tensor deformation changes the normal proper-distance
 mapping and hence the kink's coordinate profile.}
 \label{fig:wallresponse}
\end{figure*}

In the frozen regime, the wall configuration is fixed in comoving
coordinates, but its physical geometry still depends on the metric
(Fig.~\ref{fig:wallresponse}). A tensor deformation changes the proper
distance across each wall, so the local kink profile and its stress
adjust even when the wall center remains fixed. We derive this local
response and then average it over the frozen network.

Consider a locally planar wall at fixed comoving position,
$u=n_i x^i-X_0=0$, where $n_i n^i=1$ and $u$ is the
signed coordinate offset from the wall.
We calculate its stress response at fixed material configuration
to leading order in spatial gradients.\footnote{For a planar
comoving wall with vanishing initial displacement and velocity,
the tensor-induced displacement contributes to the
position-averaged stress only at second order in spatial
gradients, for a fixed frozen interval and a regular
long-wavelength expansion; see Appendix~\ref{app:displacement}.}

In the thin-wall, quasistatic limit, the local scalar profile is
$\phi\simeq\phi_K(\xi)$, where $\phi_K$ is the static planar kink
between the two degenerate vacua and $\xi$ is proper distance
normal to the wall. The scale hierarchy and validity of this
local approximation are detailed in Appendix~\ref{app:covariant}.
The metric dependence of the coordinate profile follows from
$g^{ij}n_i n_j=a^{-2}(1-h_{nn})$, where
$h_{nn}=h_{ij}n^i n^j$.
Integrating $d\xi=du/|\nabla u|$ across the thin core gives
\begin{equation}
 \xi=au\left(1+\frac12h_{nn}\right),\qquad
 \phi=\phi_K(\xi).
 \label{eq:properdistance}
\end{equation}
Thus $\delta\phi=\tfrac12h_{nn}\xi_0\phi_K'(\xi_0)$, with
$\xi_0=au$ the unperturbed proper distance: a frozen center does not imply a fixed coordinate profile.
The prime on $\phi_K$ denotes a proper-distance derivative.

For a kink between degenerate vacua, with vacuum energy subtracted,
$\tfrac12(\phi_K')^2=V$~\cite{Saikawa2017}. Its rest-frame energy density profile is
$\epsilon(\xi)=(\phi_K')^2$.
Its adjusted gradient is
$\partial_i\phi=a(1+h_{nn}/2)\phi_K'(\xi)n_i$.
Substituting into Eq.~\eqref{eq:scalarstress}, and neglecting the
subleading temporal-gradient corrections to the local kink, gives
\begin{equation}
 \begin{aligned}
 T^i{}_j=\epsilon(\xi)\big[&n^i n_j-\delta^i{}_j
 +h_{nn}n^i n_j\\
 &-h^i{}_k n^k n_j\big]+O(h^2).
 \end{aligned}
 \label{eq:localfrozen}
\end{equation}
Wall-motion contributions vanish in the frozen limit. The remaining
time dependence of the local profile is adiabatic on the core
crossing time and is neglected in the thin-wall, quasistatic expansion.
The $h_{nn}n^i n_j$ term comes from the changed field gradients,
while $-h^i{}_k n^k n_j$ comes from raising the index with the
perturbed metric. The argument of $\epsilon$ changes as well.
All three effects belong to the same kink response.

Let $\sigma=\int d\xi\,\epsilon(\xi)$ be the wall tension and
$\mathcal A_c$ the unperturbed comoving wall area per volume.
The background density is $\rho_{\rm DW}=\mathcal A_c\sigma/a$,
recovering $\rho_{\rm DW}\propto a^{-1}$ for a frozen configuration
and constant microscopic tension. At fixed normal $\bm n$, the
position average sums the single-wall stress over uniformly distributed
comoving wall centers at a fixed observation point:
\begin{equation}
 \langle T^i{}_j\rangle_{X_0}(\bm n)
 =\mathcal A_c\int_{-\infty}^{\infty}dX_0\,
 T^i{}_j(\bm x;X_0,\bm n).
 \label{eq:positionaverage}
\end{equation}
Here $\mathcal A_c$ weights the integral by the wall abundance.
The change from wall position to proper distance gives
\begin{equation}
 |dX_0|=\frac{d\xi}{a}\left(1-\frac12h_{nn}\right).
 \label{eq:positionmeasure}
\end{equation}
Combining this weight with Eq.~\eqref{eq:localfrozen} yields the
linear response at fixed orientation,
\begin{equation}
 \begin{aligned}
 \delta\langle T^i{}_j\rangle_{X_0}
 =\rho_{\rm DW}\big[&\tfrac12h_{nn}n^i n_j
 -h^i{}_k n^k n_j\\
 &+\tfrac12h_{nn}\delta^i{}_j\big].
 \end{aligned}
 \label{eq:positionresponse}
\end{equation}
The factor $1/2$ in the first term therefore includes the response
of both the gradients and the profile's integration weight.

For any function $Q(\bm n)$ of the wall normal, the isotropic
orientation average is
\begin{equation}
 \langle Q\rangle_n
 =\frac{1}{4\pi}\int_0^{2\pi}d\varphi
 \int_0^\pi d\theta\,\sin\theta\,Q(\bm n),
 \label{eq:orientationmeasure}
\end{equation}
where $\theta$ and $\varphi$ are the polar and azimuthal angles of
$\bm n$ in the unperturbed Euclidean spatial coordinates. The full
average is $\langle T^i{}_j\rangle
=\langle\langle T^i{}_j\rangle_{X_0}\rangle_n$. In particular,
\begin{align}
 \langle n_i n_j\rangle_n&=\frac13\delta_{ij},\nonumber\\
 \langle n_i n_j n_k n_l\rangle_n
 &=\frac{\delta_{ij}\delta_{kl}+\delta_{ik}\delta_{jl}
 +\delta_{il}\delta_{jk}}{15}.
 \label{eq:orientationaverage}
\end{align}
The background average gives
$\langle T^i{}_j\rangle^{(0)}=-2\rho_{\rm DW}\delta^i{}_j/3$,
consistent with $w=-2/3$. In the perturbation, the first two terms of
Eq.~\eqref{eq:positionresponse} give $\rho_{\rm DW}h^i{}_j/15$
and $-\rho_{\rm DW}h^i{}_j/3$; the trace term has no TT part.
Their sum gives the metric-induced TT stress,
\begin{equation}
 \bigl[\delta\langle T^i{}_j\rangle\bigr]_A^{\rm TT}
 =-\frac4{15}\rho_{\rm DW}h_A.
 \label{eq:response}
\end{equation}
This result applies to a statistically homogeneous and isotropic
network coarse-grained on a comoving scale $\ell$ satisfying
$L_{\rm eq}\ll\ell\ll k^{-1}$.
Although the isotropic background stress has no TT component, its
averaged linear response to a tensor perturbation is nonzero. Moving
this response to the left-hand side of Eq.~\eqref{eq:einstein} gives
an effective tensor mass term $a^2m_{\rm eff}^2h_A$, with
$m_{\rm eff}^2=16\pi G\mu$. Equation~\eqref{eq:response} recovers
the physical shear modulus $\mu=4\rho_{\rm DW}/15$ for an isotropic membrane
ensemble~\cite{Battye2005}, now as the response of the scalar kink
and its averaging measure. The effective rigidity follows without
postulating an elastic constitutive relation~\cite{Bucher1999}.
Appendix~\ref{app:covariant} provides a covariant
thin-wall derivation allowing for wall motion and displacement, and
recovers this response in the frozen, long-wavelength limit.

\section{Frozen-network source and primordial inheritance}\label{sec:source}
The isotropic average determines the metric response, while each
network realization also carries a configuration-dependent stress.
We neglect products of the network's coarse-grained anisotropy with $h_A$.
Denoting its TT component evaluated on the unperturbed FRW metric by
$[T^i{}_j]_{A,0}^{\rm TT}$, the total stress is
\begin{equation}
 \bigl[T^i{}_j\bigr]_A^{\rm TT}
 =\bigl[T^i{}_j\bigr]_{A,0}^{\rm TT}
 -\frac4{15}\rho_{\rm DW}h_A.
 \label{eq:totalstress}
\end{equation}
In the frozen regime, the configuration-dependent stress retains a
fixed comoving pattern and dilutes as $a^{-1}$, just as $\rho_{\rm DW}$
does. We therefore write
$[T^i{}_j]_{A,0}^{\rm TT}=\rho_{\rm DW}\Pi_A$, where $\Pi_A$ is a
time-independent random field describing the frozen configuration.

Substituting Eq.~\eqref{eq:totalstress} into
Eq.~\eqref{eq:einstein} and using the pure-DWD Friedmann equation,
$\mathcal H^2=(8\pi G/3)a^2\rho_{\rm DW}$, gives
\begin{equation}
 h_A''+2\mathcal Hh_A'+k^2h_A
 +\frac85\mathcal H^2(h_A-h_{{\rm ref},A})=0,
 \label{eq:master}
\end{equation}
where $h_{{\rm ref},A}=15\Pi_A/4$ and
$a^2m_{\rm eff}^2=(8/5)\mathcal H^2$. The last term combines the
effective mass term with the frozen-network source. Their balance
defines a random reference field encoding the configuration that
freezes, including its intrinsic structure as well as any inherited
primordial deformation. At $h_A=h_{{\rm ref},A}$ the total wall
TT stress vanishes. Thus $h_{{\rm ref},A}$ is the network's preferred
tensor value and the equilibrium of the leading-gradient equation.
At finite $k$,
the gradient term can still drive evolution.

We first determine a statistical input for the frozen source by
extending a scaling unequal-time correlator (UETC) whose equal-time
limit is measured in RD simulations~\cite{Pen_1997,Kawasaki_2011,Blasi2026}. These simulations contain no primordial
tensor perturbation. We denote the resulting intrinsic normalized
stress by $\Pi_A^{\rm int}$, reserving $\Pi_A$ for the full source when
a primordial mode is present. This distinction lets us add
the inherited contribution after specifying the intrinsic spectrum.

We define the intrinsic normalized-stress UETC $\mathcal U_\Pi$ by
\begin{equation}
 \langle\Pi_A^{\rm int}(\bm k,\tau)\Pi_B^{\rm int}(\bm q,\tau')\rangle
 =(2\pi)^3\delta_{AB}\delta^{(3)}(\bm k+\bm q)
 \mathcal U_\Pi(k;\tau,\tau').
\end{equation}
Here $\bm k$ and $\bm q$ are comoving wavevectors, and the Dirac
delta enforces statistical homogeneity; $\delta_{AB}$ assumes
equivalent, uncorrelated polarizations.
We continue the zero-primordial-tensor RD scaling correlator into the
frozen epoch as a prescription for $\Pi_A^{\rm int}$. It describes
stress statistics, not the tensor field already generated during RD.
The normalized-stress scaling form is
\begin{equation}
 \mathcal U_\Pi^{\rm RD}(k;\tau_1,\tau_2)
 =(\tau_1\tau_2)^{3/2}
 \mathcal C_{\rm RD}(k\tau_1,k\tau_2).
 \label{eq:RDuetc}
\end{equation}
Here $\tau_1,\tau_2$ are the two sampling times and
$\mathcal C_{\rm RD}$ is the dimensionless RD correlation shape,
including unequal-time decorrelation. The prefactor reflects the
growing comoving correlation volume.
Once the comoving configuration freezes, that correlation volume
stops growing and the normalized source becomes the same random field
at every subsequent time. In the frustrated regime,
$\Pi_A^{\rm int}(\bm k,\tau\geq\tau_{\rm eq})
=\Pi_A^{\rm int}(\bm k,\tau_{\rm eq})$.
For both times in ideal frozen DWD, the UETC is time independent.
We approximate its spectrum by matching to the RD equal-time correlator:
\begin{equation}
 \begin{aligned}
 \mathcal U_\Pi^{\rm DWD}(k;\tau_1,\tau_2)&=P_\Pi(k)\\
 &\simeq\tau_{\rm eq}^3
 \mathcal C_{\rm RD}(k\tau_{\rm eq},k\tau_{\rm eq}).
 \end{aligned}
 \label{eq:DWDuetc}
\end{equation}
Here $P_\Pi(k)$ is the dimensional power spectrum of the frozen
intrinsic normalized stress. Freezing stops the growth of the comoving
correlation volume, while matching to the RD slice approximates the
structure retained through the transition.
Let $C$ denote the infrared normalization of the frozen stress spectrum,
so that $P_\Pi(k)\simeq C\tau_{\rm eq}^3$.
As the network approaches dynamical freezing, its comoving structure
becomes fixed. We therefore assume that $C$ differs from the RD infrared
normalization by less than an order of magnitude and adopt
$C\simeq7\times10^{-4}$ as a representative estimate~\cite{Blasi2026,Dankovsky_2024,ramazanov2024spectrumgravitationalwaveslonglasting}.

The UETC continuation fixes the intrinsic stress spectrum but not the
network's response to primordial geometry, since the simulations
contain no such perturbation. Primordial transfer requires including
that dependence in the full source.

To complete the source description in the presence of a primordial
inflationary tensor $h_{{\rm inf},A}$, we separate the inherited
response from the intrinsic random stress. The fate of the primordial
mode depends on how the reference follows the primordial geometry.
Evaluation on the unperturbed FRW metric does not remove dependence
already carried by the wall-surface configuration.
Thus the full normalized frozen stress $\Pi_A$ need not be statistically
independent of the primordial mode.

We define the linear inheritance coefficient through the
cross-power spectrum of the frozen reference field and
the primordial tensor.
For statistically homogeneous and isotropic fields, with
equivalent and uncorrelated polarizations, our convention is
\begin{equation}
\langle X_A(\mathbf{k})Y_B^*(\mathbf{q})\rangle
=(2\pi)^3\delta_{AB}\delta^{(3)}(\mathbf{k}-\mathbf{q})
P_{XY}(k).
\end{equation}
For $P_{\rm inf}(k)>0$, we define
\begin{equation}
R(k)\equiv
\frac{P_{{\rm ref},{\rm inf}}(k)}{P_{\rm inf}(k)}
=\frac{15}{4}
\frac{P_{\Pi,{\rm inf}}(k)}{P_{\rm inf}(k)}.
\end{equation}
The full normalized frozen stress can then be written as
\begin{equation}
\Pi_A=\frac{4}{15}R(k)h_{{\rm inf},A}+\Pi_A^{\rm int},
\qquad P_{\Pi^{\rm int},{\rm inf}}=0.
\end{equation}
Equivalently,
\begin{equation}
h_{{\rm ref},A}=R(k)h_{{\rm inf},A}+s_A,
\qquad
s_A=\frac{15}{4}\Pi_A^{\rm int},
\qquad P_{s,{\rm inf}}=0.
\label{eq:inheritance}
\end{equation}
This is a unique decomposition into a linear projection
onto the primordial mode and an orthogonal residual.
Orthogonality refers to vanishing cross-power and does not
in general imply statistical independence.

The coefficient $R(k)$ measures primordial inheritance
by the frozen reference, not tensor transfer at DWD entry.
It is not a normalized correlation coefficient and need
not lie in $[0,1]$.
A configuration that follows primordial geometry
throughout formation and freezing realizes adiabatic
inheritance $R\to1$ in the long-wavelength limit.
A reference without a primordially correlated component has
$R=0$.
The local rigidity fixes neither $R(k)$ nor its value
through formation and freezing.  We apply the RD-to-frozen UETC continuation to this residual.
Identifying the zero-primordial-tensor source with this residual is
an additional statistical assumption; the UETC supplies no inherited
contribution.

In the continuum description of Bucher and Spergel, the material
reference metric is specified at solid formation~\cite{Bucher1999}.
Here the tensor reference field is determined by the wall configuration
reached after dynamical freezing during DWD~\cite{Zeng2026}.
The network can evolve between formation and freezing, so its primordial
inheritance must be tracked through this evolution. The intrinsic stress
spectrum alone does not determine how much primordial deformation the
frozen configuration retains.

We now use Eq.~\eqref{eq:master} to determine how primordial inheritance
and intrinsic network stress jointly govern superhorizon tensor evolution
during the frozen DWD interval.

\begin{figure}[t]
 \centering
 \includegraphics[width=\columnwidth]{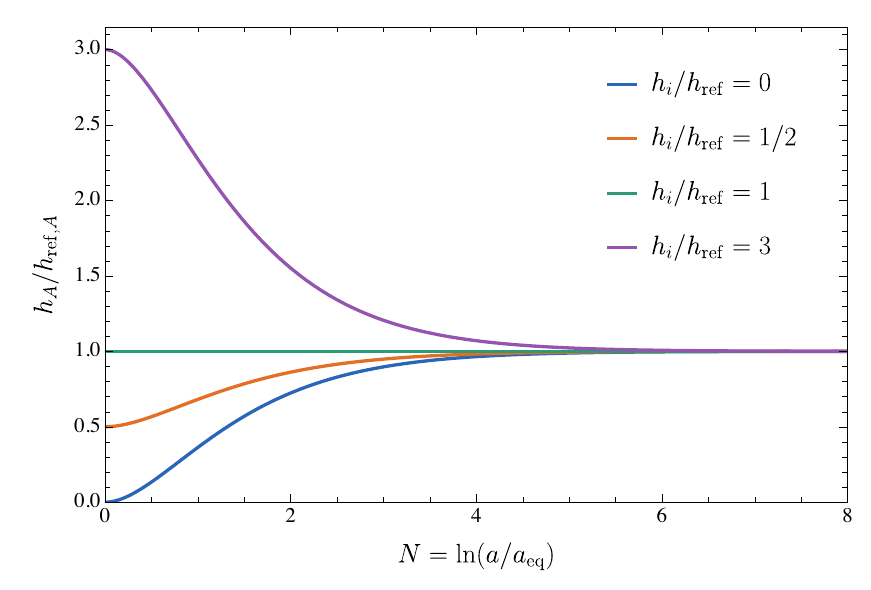}
 \caption{Analytic superhorizon relaxation toward a fixed reference
 in frozen DWD, for different initial amplitudes and vanishing initial
 tensor velocity [Eq.~\eqref{eq:replacement}].}
 \label{fig:inheritance}
\end{figure}

\section{Superhorizon tensor evolution}\label{sec:evolution}

We consider modes with $kL_{\rm eq}\ll1$ and
$k\ll k_{\rm eq}\equiv\mathcal H_{\rm eq}$, which remain superhorizon
throughout frozen DWD. Defining $N=\ln(a/a_{\rm eq})$ and
$u_A=h_A-h_{{\rm ref},A}$, Eq.~\eqref{eq:master} reduces at leading
gradient order to
\begin{equation}
 u_{A,NN}+\frac52u_{A,N}+\frac85u_A=0.
\end{equation}
Commas denote derivatives with respect to $N$. For initial amplitude
$h_{i,A}=h_A(0)$ and vanishing initial derivative $h_{A,N}(0)$,
the solution defines the relaxation function $F(N)$:
\begin{align}
 h_A(N)&=F(N)h_{i,A}+[1-F(N)]h_{{\rm ref},A},
 \label{eq:replacement}\\
 F(N)&=e^{-5N/4}\left[
 \cos\!\left(\frac{\sqrt{15}}{20}N\right)
 +\frac{25}{\sqrt{15}}\sin\!\left(\frac{\sqrt{15}}{20}N\right)
 \right].
\end{align}
Figure~\ref{fig:inheritance} illustrates this relaxation. The decaying quantity is
$h_A-h_{{\rm ref},A}=F(N)(h_{i,A}-h_{{\rm ref},A})$.
At fixed reference, the partial response to the initial amplitude
is $\partial h_A/\partial h_{i,A}=F$. This does not yet include the
reference's inherited response in Eq.~\eqref{eq:inheritance}.

Take the specified entry data $h_{i,A}=h_{{\rm inf},A}$ and
$h_{A,N}(0)=0$. Substitution of Eq.~\eqref{eq:inheritance} gives
\begin{equation}
 h_A=T_{\rm prim}(k,N)h_{{\rm inf},A}+(1-F)s_A,
 \label{eq:generaltransfer}
\end{equation}
where the primordial transfer coefficient is
\begin{equation}
 T_{\rm prim}(k,N)=F(N)+[1-F(N)]R(k).
 \label{eq:primordialtransfer}
\end{equation}
It combines the relaxation at fixed material state with the response
of that state to the primordial mode. For a sufficiently long frozen
interval, $F\to0$ and $T_{\rm prim}\to R$: the reference determines
how much primordial correlation remains after the transient decays.

For $R=1$, the reference contains the full initial tensor and
$T_{\rm prim}=1$. Its primordial-correlated component is constant,
although the total tensor can still evolve through $s_A$. This is the
adiabatic-inheritance limit at leading long-wavelength order. For
$R=0$, the reference has no primordial-correlated part and
$T_{\rm prim}=F$; the primordial coefficient is suppressed during DWD.
Partial inheritance gives an intermediate residual for real
$0<R<1$.

Equation~\eqref{eq:primordialtransfer} describes transfer within ideal
frozen DWD for the specified entry conditions. A complete cosmological
calculation must follow formation and the transition into freezing to
determine $R(k)$, the entry tensor amplitude and velocity, and their
correlations with the frozen stress, including tensors generated during
RD. The effective tensor mass alone does not fix these inputs.

With the intrinsic source spectrum and primordial transfer established,
we combine their contributions to the tensor power. We use the
dimensionless power spectrum per polarization,
$\mathcal P_X(k)=k^3P_X(k)/(2\pi^2)$.
Orthogonality in
Eq.~\eqref{eq:inheritance} gives
\begin{equation}
 \mathcal P_h=|T_{\rm prim}|^2\mathcal P_{\rm inf}
 +(1-F)^2\mathcal P_s.
 \label{eq:generalpower}
\end{equation}
The two terms describe the primordial-correlated contribution and the
contribution from intrinsic network stress, respectively.

The source spectrum is fixed by
Eq.~\eqref{eq:DWDuetc}. Since $s_A=15\Pi_A^{\rm int}/4$, its contribution
follows directly from $P_s=(225/16)P_\Pi$. In the infrared,
\begin{equation}
 \mathcal P_s(k)=\frac{225C}{32\pi^2}(k\tau_{\rm eq})^3.
 \label{eq:intrinsicpower}
\end{equation}
Its $k^3$ form follows from the finite-correlation source prescription.
This is the residual spectrum, not the full reference power when
$R\ne0$.

\begin{figure}[t]
 \centering
 \includegraphics[width=\columnwidth]{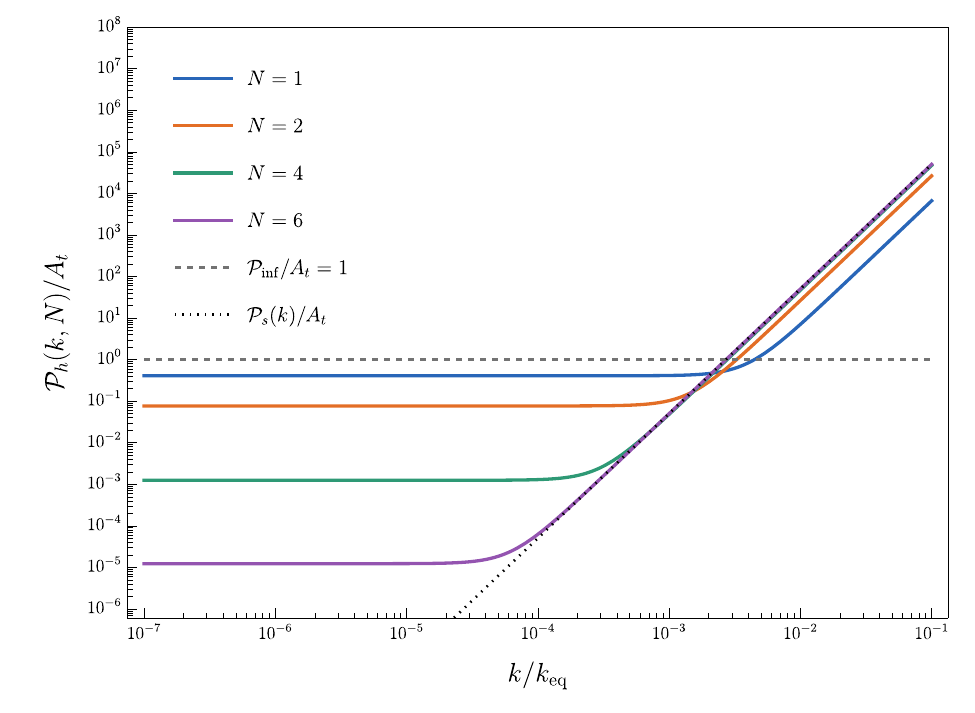}
 \caption{Illustrative tensor spectra for $R=0$ [Eq.~\eqref{eq:power}].
 Solid curves show the indicated e-folds; dashed and dotted curves denote
 the initial primordial and intrinsic reference spectra.
 Parameters are $A_t=10^{-11}$ per polarization, $C=7\times10^{-4}$,
 $k_{\rm eq}\tau_{\rm eq}=1$, and $L_{\rm eq}=k_{\rm eq}^{-1}$.
 Only the long-wavelength range $k/k_{\rm eq}\leq0.1$ is shown.}
 \label{fig:intrinsicspectra}
\end{figure}

For illustration, consider a reference uncorrelated with the primordial tensor, $R=0$, and nearly
scale-invariant primordial power $\mathcal P_{\rm inf}\simeq A_t$
per polarization. Equation~\eqref{eq:generalpower} reduces to
\begin{equation}
 \mathcal P_h(k,N)=F^2(N)A_t
 +[1-F(N)]^2\frac{225C}{32\pi^2}(k\tau_{\rm eq})^3.
 \label{eq:power}
\end{equation}
In this $R=0$ example, the primordial contribution retains its spectral
shape while its amplitude decays. The intrinsic network contribution
approaches the $k^3$ spectrum in Eq.~\eqref{eq:intrinsicpower} and can
dominate at larger $k$ within the long-wavelength regime, as shown in
Fig.~\ref{fig:intrinsicspectra}. For nonzero inheritance, the primordial
contribution instead evolves with $|T_{\rm prim}|^2$, as given by
Eq.~\eqref{eq:generalpower}.

In the absence of primordial tensor inheritance ($R=0$), the small-$k$
plateau in Fig.~\ref{fig:intrinsicspectra} decreases as $F^2(N)$,
while the rising contribution at larger $k$ grows toward the intrinsic
network spectrum. Over the interval shown, the crossover between the
two moves to smaller $k$, so that network power dominates an increasing
part of the displayed long-wavelength range. For sufficiently large $N$,
the curves approach the same $k^3$ spectrum wherever the primordial
contribution has become negligible. By contrast, adiabatic inheritance
($R=1$) would preserve the primordial plateau while allowing the intrinsic
contribution to grow.

\section{Discussion}\label{sec:discussion}
We have studied the superhorizon evolution of primordial tensor
perturbations during a dynamically frozen domain-wall-dominated epoch.
Starting from the wall-forming scalar field, we derived an effective
tensor mass term and a source determined by the frozen network
configuration. Together they drive tensor perturbations toward a state
containing both inherited primordial deformation and intrinsic network
structure. The surviving primordial component therefore depends on the
network's inheritance, rather than on the effective mass alone.

The primordial transfer is quantified by
$T_{\rm prim}=F+(1-F)R$. Adiabatic inheritance preserves the primordial
constant component even as an intrinsic network component evolves;
a non-inheriting reference suppresses the primordial coefficient.
The latter is a conditional initial-state choice, not a general
consequence of freezing. RD simulations without primordial tensors
can constrain intrinsic stress statistics, while selecting $R(k)$
requires consistent formation and initial-state dynamics. This
separation is needed when applying an effective tensor mass term to
cosmological primordial data.

Suppression of superhorizon primordial tensor modes could modify the
observed relation between the tensor-to-scalar ratio and the tensor tilt,
potentially affecting inflationary consistency tests. Assessing this
possibility requires the scalar evolution and the transition out of DWD
as well as the tensor evolution during the frozen epoch. Let $T_t$ and
$T_\zeta$ denote the complete amplitude transfers of primordial tensors
and scalar curvature perturbations, respectively. If additional sources
are negligible, the final tensor-to-scalar ratio
$r_f$ and tensor tilt $n_{t,f}=d\ln\mathcal P_{t,f}/d\ln k$ obey
\begin{equation}
 \begin{aligned}
 r_f&=\frac{|T_t|^2}{|T_\zeta|^2}r_{\rm inf},\\
 n_{t,f}&=n_{t,\rm inf}
       +\frac{d\ln|T_t|^2}{d\ln k}.
 \end{aligned}
 \label{eq:consistencytransfer}
\end{equation}
Here $r_{\rm inf}$ and $n_{t,\rm inf}$ are their inflationary values.
Tensor suppression alone therefore does not establish a change in
the single-field slow-roll relation $r_{\rm inf}=-8n_{t,\rm inf}$~\cite{Guzzetti2016}:
scalar transfer may compensate it, and scale-dependent inheritance
can also change the tilt. Our frozen-interval $T_{\rm prim}$ is an
ingredient of $T_t$, not a calculation of that complete transfer.

The result assumes thin, locally planar walls with quasistatic
profiles, vanishing initial wall displacement and velocity,
and isotropic averaging with $kL_{\rm eq}\ll1$. Modes must remain superhorizon during the
frozen interval. Within these assumptions, the effective mass and
the inherited source jointly determine primordial transfer.

These results motivate simulations that follow the wall network through
domination, dynamical freezing, and decay, with the cosmic expansion
evolved self-consistently. Including an initial primordial tensor
perturbation would allow such simulations to determine its inheritance
by the frozen network and distinguish this response from intrinsic
network stress. Following the scalar perturbations and the exit from
DWD would further establish whether the resulting tensor transfer
modifies inflationary consistency tests.

\section*{Acknowledgments}
% ======================================================================
The author thanks Sabir Ramazanov for useful discussions and valuable comments on the manuscript. The author acknowledges the support from the Foundation for the Advancement of Theoretical Physics and Mathematics “BASIS”.

\bibliography{references}

\clearpage
\appendix
\section{Covariant scalar-wall stress and its variation}
\label{app:covariant}
\subsection{Assumptions and scale hierarchy}

We use $c=1$, signature $(-,+,+,+)$, cosmic time $t$,
and conformal time $\tau$, with $dt=a\,d\tau$,
$H=\dot a/a$, and $\mathcal H=aH$.
Dots and primes denote $\partial_t$ and $\partial_\tau$,
respectively.
For a canonical scalar with degenerate vacua and
a stable kink,
\begin{equation}
S_\phi=-\int d^4x\sqrt{-g}
\left[
\frac12 g^{\mu\nu}\partial_\mu\phi\partial_\nu\phi
+V(\phi)
\right],
\qquad
\Box\phi-V_{,\phi}=0.
\end{equation}
We subtract the vacuum energy and take the microscopic
wall tension $\sigma$ to be constant.

We retain terms linear in the tensor perturbation
$h_{ij}$ and the wall-surface perturbation $f$,
and work at leading order in the thin-wall expansion.
The physical core width $\delta_w$ satisfies
\begin{equation}
\frac{\delta_w}{R_{\rm curv}}\ll1,
\qquad
H\delta_w\ll1,
\qquad
\frac{k\delta_w}{a}\ll1.
\end{equation}
Here $R_{\rm curv}$ is the local physical curvature
radius and $k$ is the comoving wavenumber.
The core is also assumed small compared with any
proper acceleration length.
We require weak gravity across the core,
$G\sigma\delta_w\ll1$.

Thinness alone does not ensure quasistatic adjustment.
Let $t_{\rm int}$ be the characteristic rest-frame
relaxation time of the relevant internal wall modes,
and let $T_{\rm geom}$ characterize changes in the
local wall geometry and imposed deformation.
We assume
\begin{equation}
H t_{\rm int}\ll1,
\qquad
\frac{t_{\rm int}}{T_{\rm geom}}\ll1,
\end{equation}
with negligible initial or driven internal-mode excitation.
Under these conditions, the leading local profile is
\begin{equation}
\phi(x)\simeq\phi_K(\xi(x)),
\qquad
\phi_{K,\xi\xi}=V_{,\phi}(\phi_K),
\end{equation}
where $\xi$ is Gaussian-normal proper distance from
the wall worldvolume in a local tubular neighborhood.
This is a local adiabatic approximation, not an exact
scalar-field solution for arbitrary FRW expansion,
wall curvature, or acceleration.

The expansion-induced error can be exhibited directly
for a planar wall at fixed comoving position.
Writing its signed comoving offset as $u$, use the
leading proper-distance mapping $\xi_0=au$ and the
trial profile $\phi=\phi_K(\xi_0)$.
Then
\begin{equation}
\dot\phi=H\xi_0\phi_{K,\xi},
\qquad
a^{-2}\nabla^2\phi=\phi_{K,\xi\xi}.
\end{equation}
Defining the cosmic-time scalar-equation operator by
\begin{equation}
\mathcal E_\phi
\equiv
\ddot\phi+3H\dot\phi-a^{-2}\nabla^2\phi+V_{,\phi},
\end{equation}
the static kink equation gives
\begin{equation}
\mathcal E_\phi[\phi_K(\xi_0)]
=
(\dot H+4H^2)\xi_0\phi_{K,\xi}
+H^2\xi_0^2\phi_{K,\xi\xi}.
\end{equation}
Across the core, $|\xi_0|\sim\delta_w$, so this residual
is of relative order
$(|\dot H|+H^2)\delta_w^2$
compared with the characteristic static core terms.
It is therefore $O((H\delta_w)^2)$ when
$|\dot H|=O(H^2)$, as in the ideal frozen DWD background.
The temporal-gradient contribution to the local stress
is suppressed by the same thin-core hierarchy.

The network average requires an additional separation
of spatial scales.
In the frozen regime, let $L_{\rm eq}$ be the comoving
network correlation length and $\ell$ a comoving
coarse-graining scale.
We assume
\begin{equation}
\frac{\delta_w}{a}\ll L_{\rm eq}\ll\ell\ll k^{-1},
\qquad
k\ll\mathcal H.
\end{equation}
The first hierarchy permits a thin-wall description
and averaging over many correlated network regions
within one perturbation wavelength.
The second restricts the calculation to superhorizon
modes.
These spatial conditions complement, but do not replace,
the local quasistatic conditions above.

\subsection{Wall geometry and covariant thin-wall stress}

Write
\begin{equation}
 \begin{gathered}
 ds^2=a^2[-d\tau^2+\widehat\gamma_{ij}dx^idx^j],\\
 \widehat\gamma_{ij}=\delta_{ij}+h_{ij},\qquad
 h^i{}_i=0,\quad \partial_i h^{ij}=0.
 \end{gathered}
\end{equation}
The physical spatial metric used below is
$\gamma_{ij}=a^2\widehat\gamma_{ij}$. Let the wall worldvolume be
$F(\tau,\bm x)=0$, and define
\begin{equation}
 \begin{gathered}
 q=\sqrt{\widehat\gamma^{ij}\partial_iF\partial_jF},\qquad
 s_i=\frac{\partial_iF}{q},\\
 s^i=\widehat\gamma^{ij}s_j,\qquad s_is^i=1.
 \end{gathered}
 \label{eq:wall-spatial-normal}
\end{equation}
Along a wall trajectory $X^i(\tau)$,
$\partial_\tau F+X^{i\prime}\partial_iF=0$. Its physical normal
velocity and Lorentz factor are therefore
\begin{equation}
 v=s_iX^{i\prime}=-\frac{\partial_\tau F}{q},\qquad
 \Gamma=(1-v^2)^{-1/2}.
 \label{eq:wall-normal-velocity}
\end{equation}
Normalizing the spacetime gradient gives
\begin{equation}
N_\mu=\frac{\partial_\mu F}
 {\sqrt{g^{\alpha\beta}\partial_\alpha F\partial_\beta F}}
 =a\Gamma(-v,s_i),\qquad
 N^\mu=\frac\Gamma a(v,s^i),
 \label{eq:wall-four-normal}
\end{equation}
since $(\nabla F)^2=q^2/(a^2\Gamma^2)>0$.
For a pure-tension wall tangential motion is a reparametrization.
Choosing trajectories with $X^{i\prime}=vs^i$ defines the unit
worldvolume velocity
\begin{equation}
U^\mu=\frac\Gamma a(1,vs^i),\qquad
 U_\mu=a\Gamma(-1,vs_i).
 \label{eq:wall-four-velocity}
\end{equation}
Thus $N^2=1$, $U^2=-1$, and $N\cdot U=0$.
The level set fixes $N$; $U$ additionally uses this normal-motion
choice and is not a unique material velocity determined by $F$ alone.

Varying the scalar action gives
\begin{equation}
 T^\mu{}_\nu=\partial^\mu\phi\,\partial_\nu\phi
 -\delta^\mu{}_\nu\left[\frac12(\partial\phi)^2+V\right].
\end{equation}
For the local rest-frame kink, let $\xi$ be normal proper distance.
At leading core order $\partial_\mu\phi=\phi_{K,\xi}N_\mu$ and the
static first integral is $\tfrac12\phi_{K,\xi}^2=V$. Hence
\begin{equation}
T^\mu{}_\nu=\epsilon(N^\mu N_\nu-\delta^\mu{}_\nu),
 \qquad \epsilon=\phi_{K,\xi}^2,\qquad
 \sigma=\int d\xi\,\epsilon.
 \label{eq:wall-covariant-stress}
\end{equation}
Within the core the normal is extended in Gaussian-normal coordinates;
only its wall value is needed in the thin-wall limit. Locally,
$F=[q/(a\Gamma)]_\Sigma\xi+O(\xi^2)$, so
\begin{equation}
 \epsilon\ \longrightarrow\ \sigma\delta_D(\xi)
 =\frac{\sigma q}{a\Gamma}\delta_D(F).
 \label{eq:wall-thin-weight}
\end{equation}
Here $\delta_D$ denotes the Dirac distribution. This surface identity
does not assume a global expression for $\xi$.
Substituting Eq.~\eqref{eq:wall-four-normal} gives
\begin{equation}
 T^i{}_j=\epsilon(\Gamma^2s^is_j-\delta^i{}_j),
 \qquad -T^0{}_0=\epsilon\Gamma^2.
 \label{eq:covariant-spatial-stress}
\end{equation}
These are local, unaveraged stresses; in particular $N^0N_0=-\Gamma^2v^2$
ensures positive energy. For fixed tension the covariant variation is
\begin{equation}
 \delta T^\mu{}_\nu
 =\delta\epsilon(N^\mu N_\nu-\delta^\mu{}_\nu)
 +\epsilon(\delta N^\mu N_\nu+N^\mu\delta N_\nu).
 \label{eq:wall-covariant-variation}
\end{equation}

\subsection{Metric and surface variations}

Set $F=F_0+f$, where $f=\delta F$ describes the wall response to the
imposed tensor $h_{ij}$. Variations are Eulerian, at fixed coordinates
and fixed $a$. All coefficients on the right below are evaluated at
$h=0$, $F=F_0$, with spatial indices raised by $\delta_{ij}$. Define
only the contractions and tangential projector
\begin{equation}
 h_{ss}=h_{ij}s^is^j,\qquad P^i{}_j=\delta^i{}_j-s^is_j.
\end{equation}
Using $\delta\widehat\gamma^{ij}=-h^{ij}$ in
Eq.~\eqref{eq:wall-spatial-normal},
\begin{align}
 2q\delta q&=-q^2h_{ss}+2q s^i\partial_if,\nonumber\\
 \frac{\delta q}{q}&=-\frac12h_{ss}+\frac{s^i\partial_if}{q},\\
 \delta s_i&=\frac{\partial_if}{q}-s_i\frac{\delta q}{q}\nonumber\\
 &=\frac12h_{ss}s_i+\frac{P_i{}^k\partial_kf}{q},\\
 \delta s^i&=\delta^{ij}\delta s_j-h^i{}_ks^k\nonumber\\
 &=-h^i{}_ks^k+\frac12h_{ss}s^i+\frac{P^{ik}\partial_kf}{q}.
 \label{eq:wall-normal-perturbations}
\end{align}
Thus $2s_i\delta s^i+h_{ss}=0$, as required by normalization.
Likewise, Eq.~\eqref{eq:wall-normal-velocity} gives
\begin{equation}
\begin{aligned}
\delta v&=-\frac{\partial_\tau f}{q}-v\frac{\delta q}{q}\\
 &=\frac v2h_{ss}-\frac{(\partial_\tau+vs^i\partial_i)f}{q},\\
 \delta\Gamma&=\Gamma^3v\delta v.
\end{aligned}
 \label{eq:wall-velocity-perturbations}
\end{equation}
The four-vectors consequently vary as
\begin{align}
 \delta N^0&=\frac{\Gamma^3}{a}\delta v,
 &\delta N^i&=\frac1a(\Gamma\delta s^i+\Gamma^3v\delta v\,s^i),\\
 \delta U^0&=\frac{\Gamma^3v}{a}\delta v,
 &\delta U^i&=\frac1a(\Gamma v\delta s^i+\Gamma^3\delta v\,s^i).
\end{align}
Covariant variations obey
$\delta N_\mu=g^{(0)}_{\mu\nu}\delta N^\nu+\delta g_{\mu\nu}N^\nu$
and similarly for $U_\mu$, with $\delta g_{ij}=a^2h_{ij}$ and
$\delta g_{0\mu}=0$.

The thin-wall weight also changes. Varying
Eq.~\eqref{eq:wall-thin-weight} at fixed $\sigma$ yields
\begin{equation}
\delta\epsilon
 =\epsilon\left(\frac{\delta q}{q}-\frac{\delta\Gamma}{\Gamma}\right)
 +\frac{\sigma q}{a\Gamma}f\delta_D'(F_0),
 \qquad \epsilon=\frac{\sigma q}{a\Gamma}\delta_D(F_0).
 \label{eq:wall-weight-perturbation}
\end{equation}
The derivative of the Dirac distribution accounts for displacement
of the wall support. In particular $f$ is a level-set perturbation;
the fixed-time coordinate-normal displacement is $-f/q$.
Combining these variations first gives
\begin{equation*}
 \begin{aligned}
 \delta T^i{}_j={}&\delta\epsilon
 (\Gamma^2s^is_j-\delta^i{}_j)
 +2\epsilon\Gamma\delta\Gamma\,s^is_j\\
 &+\epsilon\Gamma^2\left[
 h_{ss}s^is_j-h^i{}_ks^ks_j
 +\frac{P^{ik}s_j+s^iP_j{}^k}{q}\partial_kf\right].
 \end{aligned}
\end{equation*}
Substituting Eqs.~\eqref{eq:wall-velocity-perturbations} and
\eqref{eq:wall-weight-perturbation}, expanding $P^i{}_j$, and using
$\Gamma^{-2}+v^2=1$, all perturbations reduce explicitly to $h_{ij}$
and $f$:
\begin{equation}
\begin{aligned}
 \delta T^i{}_j={}&\frac{\sigma q\Gamma}{a}\Bigg\{
 \delta_D(F_0)\Bigg[
 \frac{\Gamma^2}{2}h_{lm}s^ls^m s^is_j\\
 &-h^i{}_ks^ks_j+\frac12h_{lm}s^ls^m\delta^i{}_j\\
 &+\frac{s_j\partial^if+s^i\partial_jf}{q}\\
 &-\frac{\Gamma^2s^is_j+\delta^i{}_j}{q}
       (v\partial_\tau+s^k\partial_k)f\Bigg]\\
 &+f\delta_D'(F_0)
       \left(s^is_j-\Gamma^{-2}\delta^i{}_j\right)\Bigg\}.
\end{aligned}
\label{eq:covariant-spatial-variation}
\end{equation}
Here $\partial^if=\delta^{ik}\partial_kf$, and $q,s_i,v,\Gamma,F_0$
are background quantities. The first three terms are the explicit
metric response; the remaining terms describe the displacement and
motion of the wall, including the shift of its distributional support.
The operator $v\partial_\tau+s^k\partial_k$ arises after collecting
these terms and acts only on $f$. Determining $f[h]$ requires the wall
equation of motion and initial data; no averaging has been performed.

In the frozen, fixed-surface limit used in the main text, set $v=0$
and $f=0$. For $F_0=n_ix^i-X_0$, the background has $q=1$,
$s_i=n_i$, and $\Gamma=1$, so
Eq.~\eqref{eq:covariant-spatial-variation} reduces to
\begin{equation}
 \begin{aligned}
 \delta T^i{}_j={}&\frac{\sigma}{a}\delta_D(n_kx^k-X_0)\\
 &\times\left[\frac12h_{nn}n^in_j-h^i{}_kn^kn_j
 +\frac12h_{nn}\delta^i{}_j\right].
 \end{aligned}
\end{equation}
This is the thin-wall form of the fixed-orientation response in
Eq.~\eqref{eq:positionresponse}. The position and isotropic orientation averages performed
there give $\delta\langle T^i{}_j\rangle^{\rm TT}
=-(4/15)\rho_{\rm DW}h^i{}_j$.

\subsection{Tensor-induced wall displacement}
\label{app:displacement}

The leading wall equation follows from the scalar
equation projected onto the kink's translational
zero mode. In Gaussian-normal coordinates,
\begin{equation}
\Box\phi_K-V_{,\phi}
=K\phi_{K,\xi},
\qquad K=\nabla_\mu N^\mu,
\end{equation}
where the static kink equation has been used.
At leading thin-wall order, integration against
$\phi_{K,\xi}$ gives $\sigma K|_\Sigma=0$.
Using $N^\mu=\Gamma(v,s^i)/a$, this becomes
\begin{equation}
\left[
\partial_\tau(a^3\sqrt{\widehat\gamma}\,\Gamma v)
+\partial_i(a^3\sqrt{\widehat\gamma}\,\Gamma s^i)
\right]_{F=0}=0.
\end{equation}

At $h_{ij}=0$, define
\begin{equation}
E_0=D_vv+(1-v^2)(3\mathcal H v+\partial_i s^i),
\qquad D_v=\partial_\tau+vs^i\partial_i.
\end{equation}
Then $K_0=\Gamma^3E_0/a$ and $E_0|_{F_0=0}=0$.
For a TT perturbation, the Eulerian variation is
\begin{equation}
\begin{aligned}
\delta E={}&D_v\delta v
+\left[s^i\partial_i v+3\mathcal H(1-3v^2)
-2v\partial_i s^i\right]\delta v\\
&+v\,\delta s^i\partial_i v
+(1-v^2)\partial_i\delta s^i,
\end{aligned}
\end{equation}
with
\begin{equation}
\delta v=\frac v2h_{ss}-\frac{D_vf}{q},
\qquad
\delta s^i=-h^i{}_js^j+\frac12h_{ss}s^i
+\frac{P^{ij}}q\partial_jf.
\end{equation}
The fixed-time surface displacement is
$\delta X^i=-fs^i/q$, so the linearized equation is
\begin{equation}
\left[\delta E-\frac f q s^i\partial_iE_0\right]_{F_0=0}=0.
\label{eq:wall-linearized-motion}
\end{equation}
The second term accounts for evaluating the equation
on the displaced surface.

For a planar comoving background
$F_0=n_ix^i-X_0$, with constant $n_i$, one has
$q=1$, $v=0$, and $E_0=0$ throughout the planar
extension. The TT condition then gives
\begin{equation}
f''+3\mathcal H f'-\nabla_\parallel^2f
=\frac12\partial_n h_{nn},
\label{eq:planar-displacement}
\end{equation}

where $\partial_n=n^i\partial_i$ and
$\nabla_\parallel^2=(\delta^{ij}-n^in^j)\partial_i\partial_j$.
With $f(\tau_i)=f'(\tau_i)=0$, the leading-gradient
solution is
\begin{equation}
f(\tau)=\frac12\int_{\tau_i}^{\tau}
\frac{d\eta}{a^3(\eta)}
\int_{\tau_i}^{\eta}d\eta'\,a^3(\eta')
\partial_n h_{nn}(\eta')
+\text{higher-gradient terms}.
\end{equation}
Thus a spatially uniform tensor induces no displacement,
and the induced $f$ starts at one spatial gradient.

In the homogeneous planar position average, including
the $f\delta_D'(F_0)$ term by integration by parts,
the displacement stress is
\begin{equation}
\delta_f\langle T^i{}_j\rangle_{X_0}
=\rho_{\rm DW}
\left[n_j\partial^if+n^i\partial_jf
-2n^in_j\partial_nf\right].
\end{equation}
It therefore starts at second spatial-gradient order
for a fixed frozen interval and regular long-wavelength
expansion, justifying the leading-order fixed-material
response used in the main text.
This conclusion assumes the planar background and
vanishing initial displacement and velocity; a curved
network requires the general equation
\eqref{eq:wall-linearized-motion}.

\end{document}